\RequirePackage{fix-cm}
\documentclass[smallextended]{svjour3}       
\smartqed  
\usepackage{graphicx}
\begin{document}

\journalname{Bulletin of Mathematical Biology}

\title{Reproducible research using biomodels}


\author{Pedro Mendes}


\institute{P. Mendes \at
              Center for Quantitative Medicine and Department of Cell Biology, University of Connecticut School of Medicine, Farmington, CT 06030, USA \\
              Tel.: +1-860-6797632\\
              \email{pmendes@uchc.edu}
}

\date{Received: date / Accepted: date}

\maketitle

\begin{abstract}
Like other types of computational research, modeling and simulation of biological processes (biomodels) is still largely communicated without sufficient detail to allow independent reproduction of results. But reproducibility in this area of research could easily be achieved by making use of existing resources, such as supplying models in standard formats and depositing code, models, and results in public repositories. 
\end{abstract}

\section{Introduction}
\label{intro}
Reproducibility is at the core of the scientific process. A basic aspect of science is the ability to establish reproducible effects. Karl Popper's notion of falsifiability depends on reproducibility: a theory is falsified when {\it reproducible} effects refute it \cite{popper1959}. This means that results of scientific investigations must be reproducible or else support for ``discoveries'' becomes discredited ({\it e.g.} see \cite{maddox88}). 
Therefore scientific reports must describe experiments in sufficient detail to allow other researchers to reproduce them. A recent survey suggests that there is a problem with lack of reproducibility in a large proportion of scientific articles \cite{baker2016}.

This reproducibility crisis has been highlighted mainly for experimental research. At first glance it could appear that computational research would not suffer from such problems since, after all, computers follow specific sets of instructions (programs) and thus can be run in a reproducible manner. This idea was formalized in the 1990s by Claerbout and Karrenbach \cite{claerbout1992} who described how electronic publications could easily be made into reproducible publications. For example they could allow the reader to re-execute analyses and plots directly from the data through an appropriately associated program. These authors already highlight, though, that a critical issue underlying computational reproducibility is that the programs used should be open source in order to be available to all readers. Claerbout and Karrenbach were essentially optimistic in how computational resources were going to make publications more reproducible, writing that ``{\it With workstations becoming widespread and software available, the burdens imposed on the author to create reproducible results are little more than the task of filing everything systematically}'' \cite{claerbout1992}. Unfortunately this optimism did not materialize.

It is now widely accepted that the problem with reproducibility extends also to computational research \cite{mesirov2010,peng2011,stodden2016}. And while some journals have adopted a few measures to minimize this problem  \cite{greenbaum2017,guerreiro2017,loew2015,peng2009}, a large proportion of articles describing computational research are still hard or impossible to reproduce \cite{hothorn2009,hothorn2011,hubner2011,stodden2018}.

\section{Which reproducibility?}
\label{sec:which}
It is rather unhelpful that the word `reproducibility' has been used with different meanings in this context, as summarized by Plesser \cite{plesser2018}. The terminology proposed by Goodman {\it et al.} \cite{goodman2016} seems the most appropriate and I will follow it here. These authors distinguish three different types of reproducibility:
\begin{itemize}
\item {\it reproducibility of methods} requires one to be able to exactly reproduce the results using the same methods on the same data;
\item {\it reproducibility of results} requires one to obtain similar results in an independent study applying similar procedures;
\item {\it reproducibility of inferences} requires the same conclusions to be reached in an independent replication potentially following a different methodology.
\end{itemize}
All of these types of reproducibility are desirable but should be addressed differently. The reproducibility of inferences requires that similar conclusions be reached when an independent approach is applied to a problem. Thus this aspect is perhaps the least problematic since it  can be satisfied by clearly describing the problem and the conclusions. On the other end of the spectrum, to achieve reproducibility of methods in computational research requires that the exact same software and input data be made available. To achieve reproducibility of results the methods and algorithms must be well defined and the input data is still required. Claerbout and Karrenbach were right in that to achieve all of these levels of reproducibility in computational research ``merely'' requires to describe all steps in detail --- but therein lies the problem.

\section{Biomodels}
\label{sec:biomodels}
A subset of computational research concerns the use of dynamic models of biological systems (`biomodels'). This includes a wide range of models, from those representing basic physico-chemical phenomena, such as ligand-receptor binding, all the way to models of disease transmission in populations or of entire ecosystems. Models of biochemical reaction networks or pathways are perhaps the most widely represented in this class. Computational biomodels have been reported since the dawn of computers ({\it e.g.} \cite{chance1960}) and are now increasingly used to help understand phenomena and make predictions, as can be witnessed by reading the pages of this journal and many others.
 
Research with biomodels may be better equipped to deal with reproducibility than other types of computational research. A  number of researchers have agreed on various standards and principles that promote reproducibility. Most notorious of all is the systems biology markup language (SBML, \cite{hucka2003}) a specification for biomodels that many software packages can read and write, thus promoting reproducibility (see also CellML \cite{lloyd2004}), NeuroML \cite{gleeson2010}, and PharmML \cite{swat2015}).  Standards have also been proposed for model diagrams (SBGN, \cite{le_novere2009}), simulation specifications (SED-ML, \cite{waltemath2011}), and data (SBRML, \cite{dada2010}). Finally there are also minimal information recommendations for publishing models (MIRIAM, \cite{miriam2005}) and simulations (MIASE, \cite{miase2011}) that, if followed, assure a good level of reproducibility --- essentially formalizing the process of {\it ``filing everything systematically''} advocated by Claerbout and Karrenbach \cite{claerbout1992}.

SBML allows biomodels to be specified in a manner that describes the biology and mathematics without prescribing the algorithms to be applied. An SBML file describes the transformations (``reactions'') that variables (``species'') can undergo; the kinetics of the transformations are well specified with all necessary constants, and the initial state of the system is also included. In essence an SBML file contains all that is needed for software to construct and solve the equations. Because the algorithms are not prescribed, this allows SBML models to be used in different contexts and analyzed with different formalisms in addition to those used in the original research.

With all the standards mentioned above computational results obtained with biomodels can easily be made reproducible. Publishing the biomodels as SBML files, either as attached supplementary material or by inclusion in a database like BioModels \cite{chelliah2015},  satisfies reproducibility of results because it makes the model immediately available for simulation with a range of software ({\it e.g.} COPASI \cite{hoops2006}, VCell \cite{moraru2008}, and many others). Then, by using the same software as the authors used originally, even reproducibility of methods is achieved. Because SBML does not prescribe the mathematical formalism, in some circumstances it can also facilitate reproducibility of inferences. For example, conclusions may have been obtained from an analysis using the linear noise approximation ({\it e.g.} \cite{pahle2012}), and others may reproduce the same conclusions using the Gillespie stochastic simulation algorithm \cite{gillespie1976}. This is possible because both  methods are implemented in software packages that read SBML.

\section{Publication of electronic materials}
\label{sec:publication}

The most basic aspect to promote reproducibility of research using biomodels is to publish the electronic materials used. This includes any programs used, the input data (usually constants and initial conditions), and results. Where should these materials be published? There are a number of options in current use: public repositories; journal website as supplementary materials; author's website; supplied by author upon reader's request --- listed in decreasing order of utility to the research community. 

The choice of supplying materials only when requested is not practical and there is plenty of evidence that too frequently it is not honored by authors \cite{stodden2018}; additionally those materials become inevitably lost as authors retire or die. Any results relying on materials ``supplied upon request'' should be considered non-reproducible and journal editors should simply not allow this practice. 

Publication of materials in authors' web sites is only minimally better: for a short while the materials are indeed immediately available to all, but dead links appear at a fast pace. This is due to frequent website redesigns, authors moving to other institutions, retirement, etc. 

Publication of materials in public repositories and in journal websites (as supplementary materials) are much better because the materials are immediately available and likely to be findable for a longer time. Both options have some advantages over the other and it is unclear which of the two may be better. Thus it is recommended that authors follow both whenever possible.

Several public repositories are conveniently available and are generally being used by a growing number of authors.  For code the most popular are GitHub (https://www.github.com) and CRAN (for R programs, https://cran.r-project.org/), though there are several are other options. For models the most widely used repository is the BioModels database  (https://www.ebi.ac.uk/biomodels-main/), which has the advantage of having curators that ensure the models do indeed produce the results that are described in the publications \cite{le_novere2006}. In the many cases when this is not true, they contact authors and correct the issues. Thus submission of models to BioModels already ensures a major verification of reproducibility of methods \cite{chelliah2015}. For data of any kind (which could include programs and models) other repositories could be used: Zenodo (https://zenodo.org/), Dryad (https://datadryad.org//), and FigShare (https://figshare.com/), all of which issue digital object identifiers (DOI) for data sets.

\section{Some recommendations}
\label{sec:recoomendations} 

H{\"u}bner {\it et al.} surveyed a sample of some 400 articles reporting research with biomodels and found that only a minority of them properly described the computational research performed in the study such that it could be reproduced \cite{hubner2011}. Thus it seems that despite the readily available   tools to promote reproducibility  described above, authors and journals  are not applying them widely. 
In order to improve the present situation, the list below includes actions that authors should take to make their biomodel research more reproducible. This short list is partly based on the MIRIAM proposal \cite{le_novere2009}. In addition it is also important to consult recommendations made for computational research in general \cite{piccolo2016,sandve2013,stodden2016}. 

\begin{enumerate}
\item Whenever possible use existing peer-reviewed, actively maintained and open-source software to create and analyze models. That way the algorithms and their implementation have already been reviewed, are available to all readers, and most likely will make step 3 trivial. Remember to cite the software and mention the version number used.
\item If the research required specially written software, deposit the code in a public repository, or at a minimum include it as supplementary material in the manuscript. This includes code that may require proprietary software (such as Matlab, Comsol, etc.); your programs need to be published!
\item Whenever possible, encode the model in an accepted standard (SBML, CellML, etc.),  include it as supplementary material, and submit it to a repository.
\item If it is not possible to encode the model in a standard, then include the full set of equations, parameter values, and initial conditions in the manuscript (at least in a supplement). Make sure that all algorithms used are specified unequivocally, as well as the software used (including version number). 
\item Publish the numerical results as data files, either in a repository, or as supplemental files. (Note that the generic repositories mentioned above allow very large data sets to be deposited.)
\end{enumerate}

Because some authors may disregard these recommendations, either by ignorance, for convenience of publishing quickly, or to make their research harder to reproduce, journal editors should enforce them. In particular 2, 4 and 5 are essential when 1 and 3 are not possible. As mentioned earlier, point 3 is the most comprehensive and enables all three types of reproducibility. Item 2 will only fulfill reproducibility of methods. Items 4 and 5, without any of the others, do not guarantee reproducibility but at least describe the model and results in detail.

\section{Conclusion}
\label{sec:conclusion}
Reproducibility is clearly an important aspect of science both for experimental as well as computational research. Research using biomodels should be communicated in ways that make it reproducible too. A few actions can be taken that will greatly facilitate this objective. Scientific journals should not publish non-reproducible research and thus should promote, or even enforce, such actions. 


\bibliographystyle{spmpsci}      
\bibliography{reproducibility.bib}   

\end{document}